\documentclass{PoS}
\usepackage{graphics}

\PoS{PoS(LAT2005)277}

\title{Distributing the chiral and flavour components of Dirac-K\"{a}hler fermions across multiple lattices.}

\ShortTitle{Dirac-K\"{a}hler fermions and multiple lattices.}

\author{\speaker{Steven ~Watterson}\\
        School of Mathematics, Trinity College, Dublin 2, Ireland\\
        E-mail: \email{watterss@maths.tcd.ie}}

\author{James Sexton\\
        School of Mathematics, Trinity College, Dublin 2, Ireland\\
        E-mail: \email{sexton@maths.tcd.ie}}

\abstract{We use a specific implementation of discrete differential geometry to describe Dirac-K\"{a}hler fermions in such a way that we can separate their chiral and flavour components.  The formulation introduces additional lattices so that on each lattice there is a single field of definate chirality.  Within this framework, we define an non-compact Abelian gauge theory.}

\FullConference{XXIIIrd International Symposium on Lattice Field Theory\\
		 25-30 July 2005\\
		 Trinity College, Dublin, Ireland}

\begin{document}

\section{Introduction}

When fermionic field theories are described using differential geometry the counterpart of the Dirac operator is the Dirac-K\"{a}hler operator and its solution is a linear combination of differential forms.  Attempts to discretize this description have always come unstuck because of the problems encountered defining discrete differential geometries.   The difficulty lies in constructing a wedge product and a Hodge star operator that are local and permit the exterior derivative to fulfill Leibniz' rule.  

Becher and Joos were one of the first to address this issue \cite{BJ} and they produced a formulation that is analogous to staggered fermions with a non-local wedge and Hodge star .  Subsequent attempts have been made that invariably sacrifice one of the target properties (eg \cite{KK}).  These difficulties reflect the consequences of the no-go theorem of Nielsen and Ninomiya \cite{NN} for discrete differential geometries. 

We shall tackle this problem by applying a formulation of discrete differential geometry devised by David Adams \cite{DA} to the Schwinger model in such a way that we isolate the chiral and flavour components of the fermion fields.  What makes the formulation uniquely appropriate for this application is that it introduces an extra lattice, parallel to the original.  We define the counterpart of $\gamma_5$ to map between these lattices and this enables us to project out the flavour and chiral components on to each lattice.  In some sense, we have used the problems discrete differential geometry as a tool to unpick the problem of fermion doubling. 

\section{The Dual Lattice and Discrete Differential Geometry}
The Hodge star is an operator that maps a differential form to its spacial complement.  In the continuum it is defined as $*dx^H = \epsilon_{H,\mathcal{C}H}dx^{\mathcal{C}H}$ where $dx^H$ is the ordered product $dx^{H_1} \wedge dx^{H_2}\wedge..\wedge dx^{H_h}$, the ${H_i}$ are components of the space and $h$ is the number of them in the ordered set $H$.  $\mathcal{C}H$ is the complement of $H$ in the space and $\epsilon$ is the Levi-Civita tensor. 

The problem with recreating this on the lattice is that to map from a form to one of complementary dimension, $*$ must be defined to be non-local.  This has the consequence that $**$ is not proportional the identity, but to a translation operator.  To resolve this, Adams introduced a dual lattice that overlies the original.  

\begin{minipage}[h]{5.5cm}
\unitlength1cm
\begin{picture}(5.5,6.0)
\resizebox{5cm}{5cm}{\includegraphics{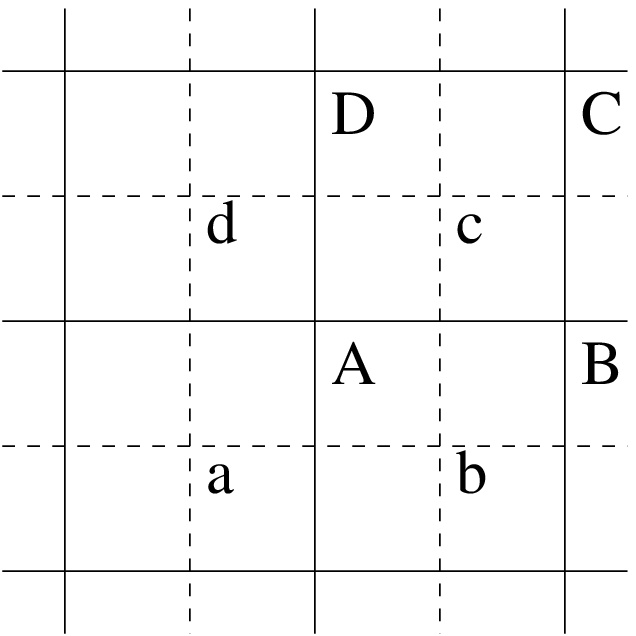}}
\end{picture}
\end{minipage}
\begin{minipage}[h]{8.8cm}
In the diagram on the left, the solid line represents the original lattice and the dashed line represents the dual lattice.  $[AB]$ is a cochain analogous to a 1-form in the $x_1$ direction.  $*$ maps it to $[bc]$ which is a cochian analagous to a 1-from in the $x_2$ direction on the dual lattice.  $*$ also maps $[AD]$ to $-[dc]$, along with $[A]$, which is analogous to a zero form, to $[abcd]$ and $[ABCD]$ to $[c]$.  Similarly, $*$ maps back from the dual to the original lattice: $*[dc]=[AD]$, $*[bc]=-[AB]$, $*[c]=[ABCD]$ and $*[abcd]=[A]$
\end{minipage}

For a local, discrete differential geometry, we use the formulation introduced by Adams and adapted to this application by Vivien de Beauc\'{e} and Samik Sen \cite{VS1} \cite{VS2} \cite{VS3}.  Continuous differential fields are discretized with the De Rahm map ({\it R}), which integrates each form of the field over the simplices of the complex with the same dimension: $\eta_1(x) dx_1$ is integrated over 1D simplices that lie parallel to the $x_1$ dimension, $\eta_{2}(x)dx_2$ is integrated over the simplices that lie parallel to the $x_2$ direction, $\eta(x)$ is sampled at the vertices of the lattice and $\eta_{12}(x)dx_1dx_2$ is integrated over the the 2D simplices.  This furnishes us with the discrete fields $\tilde{\eta}([..])$ where $[..]$ denotes the simplex used as the domain of integration.

The Whitney map ({{\it W}) acts to replace the simplicies with differential forms of the same dimension.  It also introduces linear functions in the remaining dimensions to interpolate between neighbouring simplices.

With this framework, we can define a discrete wedge product by $dx^{\mu} \wedge^{K} dx^{\nu}$ = $R[W[dx^{\mu}] \wedge W[dx^{\nu}]]$.  We can also construct a lattice derivative, $D$, from the continuum definition, $d$, as $D = RdW$.  The adjoint derivative can be constructed using both the $*$ and $D$ definitions: $\delta = (-1)^{nh+n+1}*D*$, where $h$ is the dimension of the simplex and $n$ is the dimension of the space. 

\section{Chiral \& Flavour Symmetries}
As explained in \cite{BJ}, the matrix $Z$ is used to relate the Dirac and Dirac-K\"{a}hler bases. In 2D, $\psi(x)$ is a $2*2$ matrix and $\phi(x)$ is a field comprising four differential forms.  $\phi(x) = \sum_{H} \phi(x,H) dx^H = \sum_{ab}\psi_a^{(b)}(x)Z_{ab}$.  We can explicitly relate the two fields with $\phi(x,H) = Tr(\sigma_H^{\dagger}\psi(x))$, where $\sigma$ are the Pauli matrices.

Rabin showed that the Hodge star was related to the chiral symmetry of differential fields \cite{Rabin}.  In the continuum, replacing $\psi(x)$ with $\sigma_3\psi(x)$ corresponds to applying the operation $-i*\mathcal{BA}$ to $ \phi(x)$, where $\mathcal{B}$ and $\mathcal{A}$ are operators with the properties: $\mathcal{A}dx^H=(-1)^h dx^H$ and $\mathcal{B}dx^H = (-1)^{h \choose 2}dx^H$.

In the discrete theory, we have two lattices each with their own fields.  As in \cite{VS1}, we separate the Hodge star into two operators that map from one lattice to the other: $* = *_o + *_d$.  Here we have that $*_o$ maps from the original lattice to the dual and $*_d$ maps from the dual lattice to the original.  Applying $*$ twice yields $**=*_o*_d + *_d*_o$, which is proportional to the identity.

For our formulation, we introduce a phase into the Hodge star: $*=i*_o - i*_d$, so that $**$ is still $*_o*_d + *_d*_o$, but a difference of sign is introduced in going between each lattice.

The original lattice is the physical lattice, so we put on it the fermion field discretized with the De Rham map.  However, the dual is artificial and we are free to put whatever fields on it we like. So we construct a field that results in a separation of the chiral components when we apply $P_{R/L}=\frac{1}{2}(1 \mp i*\mathcal{BA})$. 

We construct the following field for the dual: $i\tilde{\phi}([H]_d, [K]_o)$.  In this definition, a factor of $i$ is included to match the definition for $*$.  $[H]_d$ denotes the simplex of the dual lattice to which the field $\tilde{\phi}$ belongs.  The de Rahm map is not used in the construction of these fields because the domains of integration must match those of the original lattice.  So, instead, we discretize the field by integrating over the complementary simplices from the original lattice.  The domain of integration is denoted by $[K]_o$, where $*_d[H]_d = \pm [K]_o$.  Fitting the domains of integration to the original lattice ensures that the domains of the original and the dual match, so that terms can cancel when we apply $P_{R/L}$.  

With this field on the dual, application of $P_{R/L}$ projects one chiral component on to the original lattice and the other on to the dual lattice.

To separate the degenerate flavours of $\phi(x)$ we must isolate the columns of $\psi(x)$.  In the continuum, this achieved through right multiplication of $\psi(x)$ with $P_{R/L}$, which we rewrite as $P_{1/2}$ in this context.  Right mutliplication of $\psi$ by $\sigma_3$ corresponds to the operation $-i* \mathcal{B}$ on $\phi(x)$.  On our lattices, we can isolate the degenerate flavours of $\phi$ using the projection operator $P_{1/2}=\frac{1}{2}(1\mp i * \mathcal{B})$.

We now have the power to separate the chiral or flavour components of the fermion field.  However, at this stage we cannot isolate both simultaneously.  Using just the original lattice and the dual, we could apply the flavour projection operators so that flavour $1$ was on the original lattice and $2$ was on the dual, but this would eliminate the possibility of cancellation between chiral components when we applied $P_{R/L}$.

To over come this we need to introduce duplicate copies of both the original and dual lattices.  If we label the two sets of lattices by $A$ and $B$, the diagram below shows how we can separate the chiral and flavour components simultaneously. 

\begin{minipage}[h]{6.5cm}
\unitlength1cm
\begin{picture}(6.5,6.5)
\resizebox{6cm}{6cm}{\includegraphics{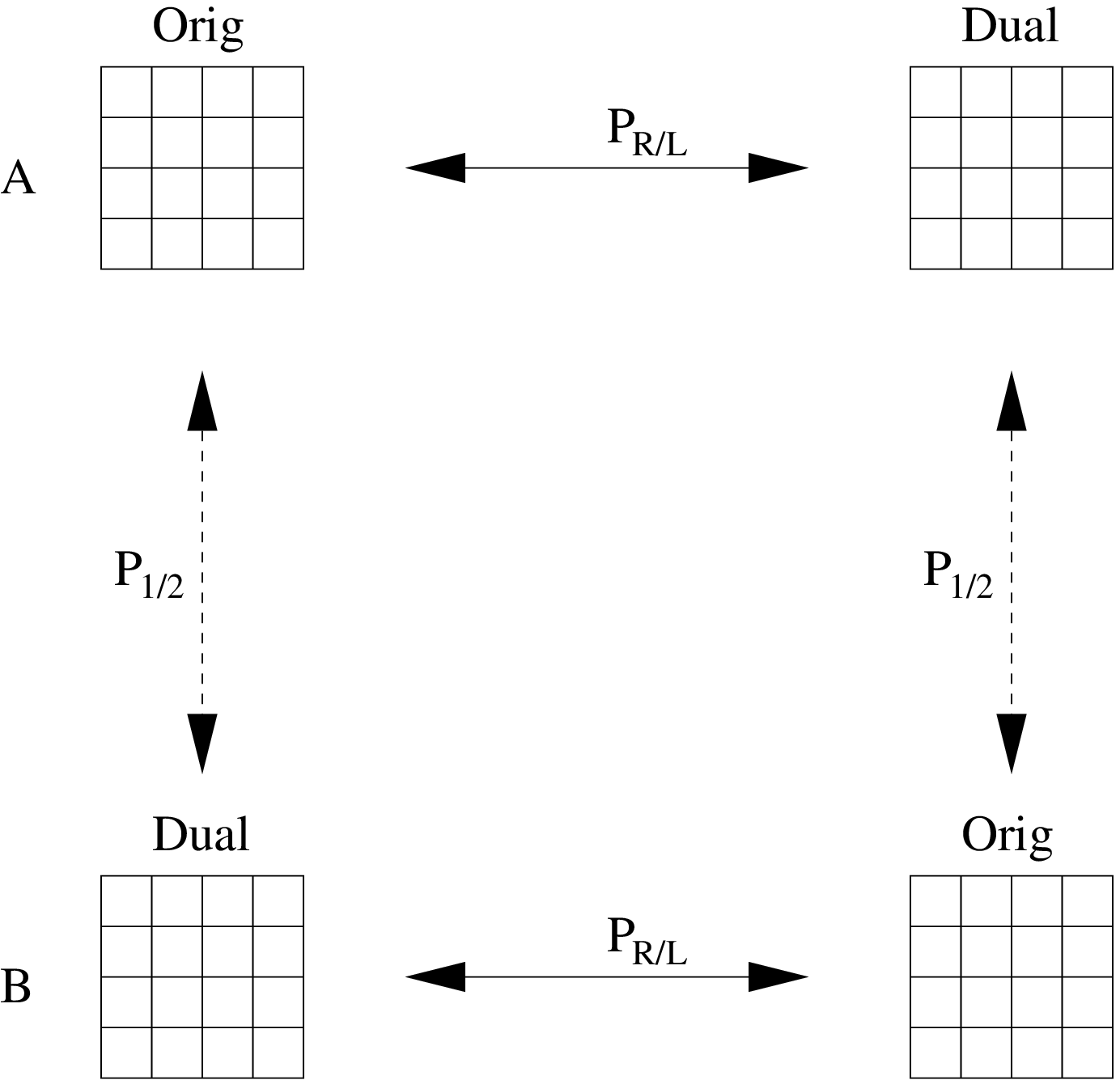}}
\end{picture}
\end{minipage}
\begin{minipage}[h]{7.8cm}
Imagine starting with two identical copies of both lattices, copy $A$ and copy $B$.  Both copies start with the same fields.  We apply the flavour operation to $A$ in such a way that flavour $1$ is projected on to the the original lattice of $A$ and flavour $2$ is projected onto the dual of $A$.  On copy $B$ we apply the flavour projection so that $2$ lies on the original lattice and $1$ lies on the dual.  We define the chiral projections operators to map between the original and dual lattices of different copies: between the original of $A$ and the dual of $B$ and vice versa.  We can now apply the chiral projectors so that the both flavours separate into their chiral components.  We put the $+$ component of flavour $1$ on the original lattice of $A$ and the $-$ component on the dual of $B$.  Similarly, we put the $+$ component of flavour $2$ on the dual of $A$ and the $-$ component on the original lattice of $B$. 
\end{minipage}

The ususual domain of integration that we have chosen for the field on the dual lattice does not impede our use of the Dirac-K\"{a}hler operator on the dual.  The four lattices independently respect the Dirac-K\"{a}hler algebra.  Furthermore, from the continuum relations, we can show that $\{-i*\mathcal{BA}, d-\delta\}=0$ and $[-i*\mathcal{B}, d-\delta]=0$, so the application of $P_{R/L}$ and $P_{1/2}$ yield solutions that also respect the Dirac-K\"{a}hler algebra.

\section{Abelian Gauge Fields}
Our next step is to add gauge fields and we start with the Abelian case.  In the continuum, the covariant Dirac-K\"{a}hler equation is $(d-\delta)\Phi(x) = ieA(x)\vee\Phi(x)$, where $\vee$ denotes the clifford product and $A(x)$ is the gauge field which is a one-form.  We discretize $A(x)$ with the De Rahm map, making the formulation a non-compact discrete fermionic field theory. Examples of other non-compact discrete field theories include \cite{MG} \cite{HN} and \cite{FP}. 

To define Abelian gauge transforms, we start with the continuum transformation $\psi(x) \rightarrow e^{i\theta(x)}\psi(x)$ which corresponds to the product between the spinor and a zero form.  In our discretization, the discrete analogues of zero forms are to be found at the vertices of the lattice and so the discrete Dirac-K\"{a}hler field transforms as
\[
\tilde{\phi} \rightarrow R\left[W\left(e^{i\tilde{\theta}}\right)W\left(\tilde{\phi}\right)\right]
\]
where the $\tilde{\theta}$ are the values of $\theta(x)$ at the vertices.  The adjoint Dirac-K\"{a}hler field transforms as 
\[
\bar{\tilde{\phi}}([..])\rightarrow \bar{\tilde{\phi}}([..])\left[R\left\{W(e^{i\tilde{\theta}}) W([..])\right\}\right]^{-1}
\]

Through Leibniz' rule, we can transform the gauge fields to keep the action invariant.  However, there are some subtleties here.  We must transform gauge fields on both lattices, which gives us
\[
\tilde{A}_o \rightarrow \tilde{A}_o -\frac{i}{g}R\left[dW(e^{i\tilde{\theta}})\right]\left[G([..])\right]^{-1}
\]

\[
\tilde{A}_d \rightarrow \tilde{A}_d -\frac{i}{g}R\left[dW(e^{i\tilde{\theta}})\right]\left[G([..])\right]^{-1}
\]
where $G[..]$ is the gauge transform on the simplex to which $A$ is being applied. However, in order that the gauge simplices remain local and gauge invariant, we must rewrite the covariant Dirac-K\"{a}hler equation slightly.  In the expression $(d-\delta)\tilde{\phi} = ie\tilde{A}\vee^K \tilde{\phi}$, we explicity write $\delta$ as $-*D*$.  Now we can see that the gauge field on the original lattice must transform so as keep the fermion field on original invariant after $D$ is applied.  Similarly, the gauge field on the dual must transform so as to keep the fields invariant after $-*D*$ is applied.  To achieve this in such a way that that is local, we must replace $A\vee^K$ with $A_o \wedge - *A_d\wedge*$ where $A_o$ and $A_d$ are the fields on the original and dual lattices, respectively.  We must also introduce a transformation law for the Hodge star. 
\[
\mbox{For } *[M]=\pm[K]\hspace{0.5cm}:\hspace{0.5cm}  * \rightarrow G([H])*\left[G([M])\right]^{-1}
\]

\section{Gauge Action}
With this description, we can define the gauge action and Lorentz gauge fixing terms as 
\[
S_G = -\frac{1}{2}<D\tilde{A}, D \tilde{A}> \hspace{1cm} S_{GF}=-\frac{1}{2\xi}<\delta \tilde{A}, \delta \tilde{A}>
\]
where the gauge invariance of the former follows from the identity $D^2=0$.

\section{Conclusion}
In the preceding sections, we saw that by interpretting the Hodge star as a map between separate lattices we could isolate the chiral and flavour components of solutions to the 2D Dirac-K\"{a}hler equation.  This necessitated the use of four lattices.  We also constructed an Abelian field theory on these lattices complete with action and gauge fixing terms.  The separate contributions from the gauge field on each lattice meant that we had to construct the transformation laws carefully, modifying subtly the definition of the covariant Dirac-K\"{a}hler equation.  In subsequent publications, we hope to extend this work further.   

SW acknowledges his gratitude for support from Enterprise Ireland.

\end{document}